\documentclass[a4paper,fleqn]{cas-dc}
\usepackage[justification=centering]{caption}
\usepackage{bm}
\usepackage[authoryear,longnamesfirst]{natbib}
\usepackage[longnamesfirst, authoryear]{natbib}  
\usepackage[justification=centering]{caption}
\usepackage[mathscr]{eucal}
\usepackage{hyperref}

\def\tsc#1{\csdef{#1}{\textsc{\lowercase{#1}}\xspace}}
\tsc{WGM}
\tsc{QE}

\begin{document}
\let\WriteBookmarks\relax
\def\floatpagepagefraction{1}
\def\textpagefraction{.001}

\shorttitle{\title{CMATH: Cross-Modality Augmented Transformer with Hierarchical Variational Distillation for Multimodal Emotion Recognition in Conversation}}
\shortauthors{Zhu et~al.}

\title[mode = title]{CMATH: Cross-Modality Augmented Transformer with Hierarchical Variational Distillation for Multimodal Emotion Recognition in Conversation}




\author[1]{\textcolor{black}{Xiaofei Zhu}}[orcid=0000-0001-8239-7176]
\ead{zxf@cqut.edu.cn}
\credit{Conceptualization of this study, Methodology, Writing - original draft \& review \& editing, Supervision}
\cormark[1] 

\author[1]{\textcolor{black}{Jiawei Cheng}}
\ead{chengjiawei@stu.cqut.edu.cn}
\credit{Conceptualization of this study, Methodology, Software, Writing - original draft}

\author[2]{\textcolor{black}{Zhou Yang}}
\ead{200310007@fzu.edu.cn}
\credit{Supervision}

\author[1]{\textcolor{black}{Zhuo Chen}}
\ead{chenzhuo@cqut.edu.cn}
\credit{Supervision} 

\author[3]{\textcolor{black}{Qingyang Wang}}
\ead{wangqingyang@caeri.com.cn}
\credit{Supervision} 

\author[4]{\textcolor{black}{Jianfeng Yao}}
\ead{yaojf@changan.com.cn}
\credit{Supervision} 

\address[1]{College of Computer Science and Engineering, Chongqing University of Technology, Chongqing 400054, China} 
\address[2]{College of Computer and Data Science, Fuzhou University, Fuzhou 350108, China} 
\address[3]{Wind Tunnel Center, China Automotive Engineering Research Institute Co., Ltd., Chongqing 401120, China}
\address[4]{Big Data Center, Chongqing Changan Automobile Co., Ltd., Chongqing 400023, China}
\cortext[cor1]{Corresponding author}

\begin{abstract}
Multimodal emotion recognition in conversation (MER) aims to accurately identify emotions in conversational utterances by integrating multimodal information. Previous methods usually treat multimodal information as equal quality and employ symmetric architectures to conduct multimodal fusion. However, in reality, the quality of different modalities usually varies considerably, and 
utilizing a symmetric architecture is difficult to accurately recognize conversational emotions when dealing with uneven modal information. Furthermore, fusing multi-modality information in a single granularity may fail to adequately integrate modal information, exacerbating the inaccuracy in emotion recognition. In this paper, we propose a novel \underline{C}ross-\underline{M}odality \underline{A}ugmented \underline{T}ransformer with \underline{H}ierarchical Variational Distillation, called CMATH, which consists of two major components, i.e., Multimodal Interaction Fusion and Hierarchical Variational Distillation. The former is comprised of two submodules, including Modality Reconstruction and Cross-Modality Augmented Transformer (CMA-Transformer), where Modality Reconstruction focuses on obtaining high-quality compressed representation of each modality, and CMA-Transformer adopts an asymmetric fusion strategy which treats one modality as the central modality and takes others as
auxiliary modalities. 
The latter first designs a variational fusion network to fuse the fine-grained representations learned by CMA-Transformer   into a coarse-grained representations. Then,  it introduces a hierarchical distillation framework to maintain the consistency between modality representations with different granularities.
Experiments on the IEMOCAP and MELD datasets demonstrate that our proposed model outperforms previous state-of-the-art baselines. Implementation codes can be available at \url{https://github.com/cjw-MER/CMATH}.
\end{abstract}


\maketitle

\section{Introduction}

Emotion Recognition in Conversation (ERC) aims to identify the corresponding emotion of each utterance in a conversation. It has become a critical problem in recent years due to its potential  applications in social media analysis \citep{kumar2015emotion}, recommendation systems \citep{xu2018exploiting,zheng2022perd}, healthcare services \citep{pujol2019emotion}, and affective computing systems \citep{zhou2020design}.  

Early research works usually assume that the emotion of the current utterance depends on its surrounding utterances as well as the speaker's emotional state. DialogueRNN \citep{majumder2019dialoguernn} presents a RNN-based neural network to keep track of individual party states. DialogueGCN \citep{ghosal2019dialoguegcn} models inter- and self-party dependency with a graph convolutional network  to capture emotional inertia of individual speakers. Although these methods demonstrates impressive performance in emotion recognition, they only focus on unique modality (i.e., textual modality).  

In recent years, multimodal emotion recognition in conversation has attracted considerable attention which attempts to model emotions of utterances by utilizing multiple modality information, such as textual, audio, and visual cues.  Recent works mainly employ transformer mechanisms \citep{lian2021ctnet,zou2023mpt-hcl} or interaction graphs \citep{hu2021mmgcn,hu2022mmdfn,nguyen2024curriculum,tu2024adaptive} to model interactive relationships across different modalities \citep{ma2024transformer}.
Lian et al. (\citeyear{lian2021ctnet}) apply a cross-modal transformer to capture intra- and inter-modal interactions. 
Zou et al. (\citeyear{zou2023mpt-hcl}) introduce a prompt transformer for cross-modal information interaction.
Hu et al. (\citeyear{hu2021mmgcn}) propose a multimodal fused graph convolutional network to explore both multimodal and long-distance contextual information. Hu et al. \citep{hu2022mmdfn} design a graph-based dynamic fusion paradigm to reduce redundancy and boost complementarity between different modalities. 
Nguyen et al. (\citeyear{nguyen2024curriculum}) employ a directed acyclic graph to integrate multimodal information. 
Tu et al. (\citeyear{tu2024adaptive})  develop an adaptive interactive graph network  to enhance intra- and cross-modal interactions.

Despite previous works have demonstrated promising results, they  still suffer from several limitations:
\begin{figure}
  \centering
  \includegraphics[width=0.4\textwidth]{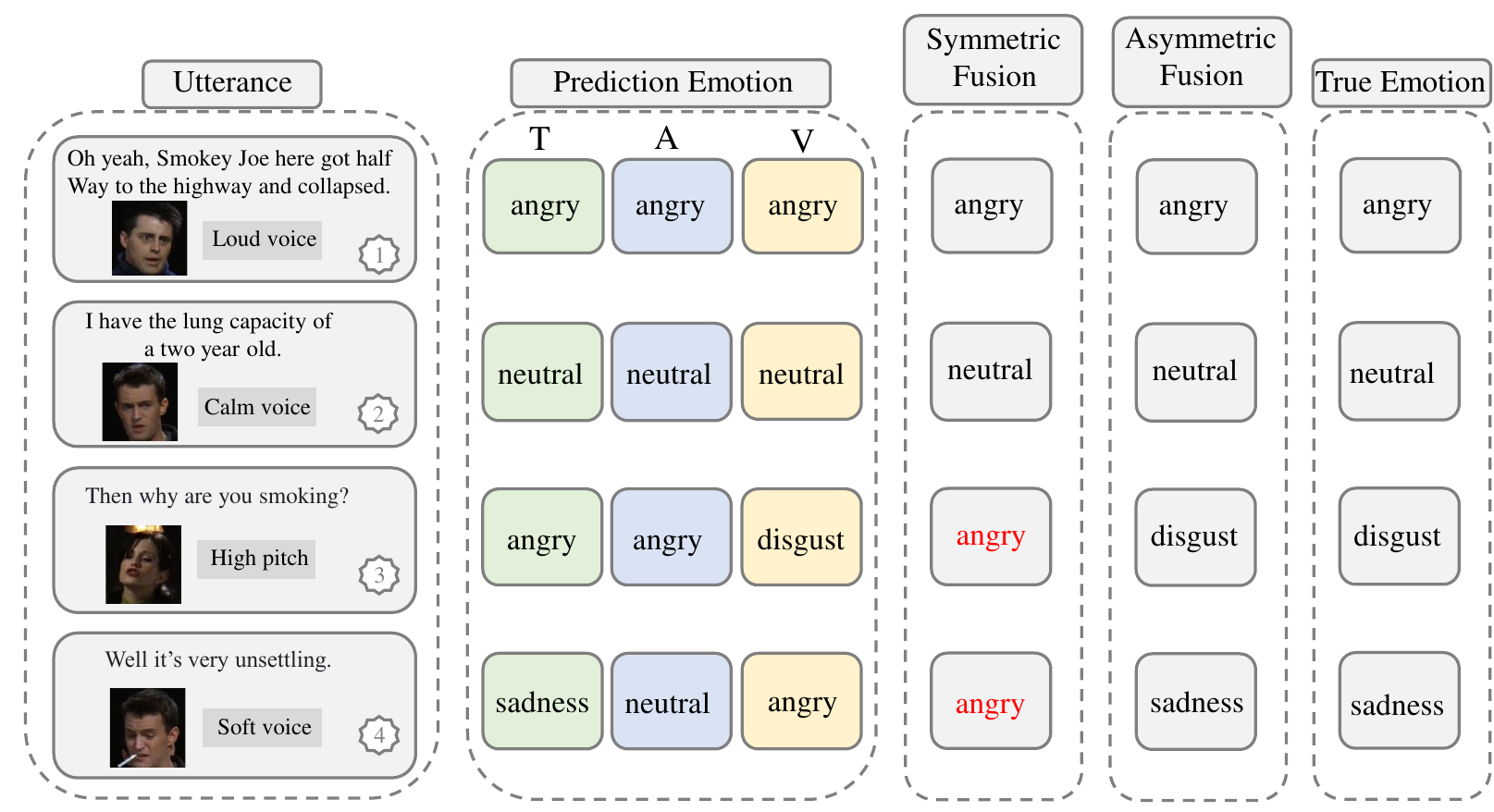}
  \caption{An example of a multimodal conversation scenario, in which the conversational contents are from textual, audio and visual modalities. }
  \label{fig: example}
\end{figure} 
Firstly, they usually treat different modalities as equivalent quality and  employ symmetric architectures to fuse information in these modalities. 
However, in reality, modality qualities would be various, and the symmetric fusion strategy inevitable leads to inferior performance.
Secondly, existing fusion methods mainly focus on fusing multimodal information in a unique information granularity by employing a unique-stage fusion strategy. They ignore the rich multi-granularity information embedded in different modalities, including the fine-grained feature representation and the coarse-grained semantic representation.


To address the aforementioned issues, this paper proposes a novel \underline{C}ross-\underline{M}odality \underline{A}ugmented \underline{T}ransformer with \underline{H}ierarchical Variational Distillation (CMATH). It consists of two major components: Multimodal Interaction Fusion module and Hierarchical Variational Distillation module. In the Multimodal Interaction Fusion module, we introduce a cross-modality augmented transformer (CMA-Transformer)  for fine-grained feature-fusion. Different from previous symmetric fusion strategy, CMA-Transformer adopts an asymmetric fusion strategy, which treats one modality as  the central modality  and the others  as auxiliary ones. To be specific, it attempts to enhance the quality of the central modality by exploring interaction information from these auxiliary modalities. We employ CMA-Transformer for each modality by taking it as the  central modality, and obtain its augmented representation. 
In the Hierarchical Variational Distillation Module, we first introduce a   variational fusion network for coarse-grained semantic information fusion, which takes the above augmented fine-grained feature representation for each modality as input and learns a multimodal Gaussian distribution.
Then, we introduce a Hierarchical Distillation strategy to maintain the consistency between modality representations with different granularities to further enhance the effectiveness of the semantic distribution integration.

To validate the effectiveness of the proposed model CMATH, we conducted extensive experiments on two widely used datasest, i.e., IEMOCAP and MELD.
Experimental results demonstrate that CMATH significantly outperforms all baseline methods and achieves the state-of-the-art performance. Specifically, the  proposed method obtains 73.90\% and 73.96\% in terms of two metrices ( i.e., accuracy and weighted F1-score) on the dataset IEMOCAP, with the relative performance improvement of 4.84\% and 4.55\% over the best performance baseline AdalGN, respectively. We further carry out experiments to analyze the effectiveness of each component of CMATH in depth in order to explore their contribution to the performance of multimodal emotion recognition in conversation.
In summary, the main contributions of this work are as follows: 
\begin{itemize}
\item We propose to treat different modalities as nonequivalent quality during the fusing process and propose  a novel CMA-Transformer which fuses multi-modality in an asymmetric strategy. We focus on enhancing the quality of each modality by taking it as the central modality and  exploiting the remaining modalities as auxiliary information.
\item We introduce a hierarchical variational distillation framework, which first leverages a light variational fusion module to extract coarse-grained semantic information by fusing the Gaussion distribution of each modality, and then applies a hierarchical distillation strategy to maintain the consistency between modality representations with different information granularities.
\item We validate the effectiveness of our proposed method on two widely used benchmark datasets, i.e., IEMOCAP and MELD. Experimental results 
demonstrate its superiority over existing state-of-the-art methods.
\end{itemize}

\section{Related Work}   
Previous research efforts \citep{majumder2019dialoguernn, ghosal2019dialoguegcn, ma2022mvn} on emotion recognition in conversation (ERC) mainly focus on unique modality, i.e., textual modality.  
Majumder et al. \citeyearpar{majumder2019dialoguernn}  propose an attentive RNN called DialogueRNN which employs three GRUs, i.e., global GRU, party GRU and emotion GRU, to model the speaker, the preceding context  and the preceding emotion.  
Ghosal et al. \citeyearpar{ghosal2019dialoguegcn} extend DialogueRNN by proposing  DialogueGCN which explores the emotional inertia of each speaker via capturing both intra- and inter-speaker dependencies. Moreover, it incorporates the relative position of context utterances from the target utterance to model the influence of both past and future utterances.
Ma et al. \citeyearpar{ma2022mvn} develop MVN which explores the emotion representation of the query utterance based on two dependencies, i.e., the word-level dependencies among utterances and the utterance-level dependencies in the context.  

Since a unique modality would  provide insufficient information as compared to multimodal for ERC, recently there has been growing interest in multimodal emotion recognition in conversation. 
Despite an early concatenation operation can be applied for capturing multimodal features for the above-mentioned methods, they neglect the rich interaction information across different modalities and result in suboptimal results. 
To address the issue, 
Lian et al. \citeyearpar{lian2021ctnet} utilize the transformer-based structure to capture intra- and inter-modal interactions among different modalities, and model temporal information in the utterance by considering both word-level lexical features and segment-level acoustic features.
Hu et al. \citeyearpar{hu2021mmgcn} propose a multimodal fused graph convolutional network named MMGCN by  creating a graph based on all modalities. It builds the connected graph in each modality and constructs edge connections between nodes across different modalities. 
Hu et al. \citeyearpar{hu2022mmdfn} leverage graph convolution operation to aggregate dynamics of contextual information of both intra- and inter-modality in a specific semantic space, and  capture the intrinsic sequential patterns of contextual information in adjacent semantic space.  By control information flow between layers, this method can reduce redundancy and boost  complementarity between modalities.
Yang et al. \citeyearpar{yang2023scmm} propose the self-adaptive context and modal-interaction modeling framework, which attempts to model different ranges of context dependency  as well as captures the specific contribution of each modality.  
Zou et al. \citeyearpar{zou2023mpt-hcl} leverages audio and visual modalities as the prompt information for textual modality and designs a multimodal prompt transformer for cross-modal information interaction. Then a hybrid contrastive learning strategy is incorporated to improve the performance for less sample labels.
Nguyen et al. \citeyearpar{nguyen2024curriculum} attempt to integrate multimodal features by employing Directed Acyclic Graph (DAG), and introduce the curriculum learning strategy to handle emotional shifts and  data  imbalance.
Tu et al. \citeyearpar{tu2024adaptive} proposes an adaptive interactive graph network called AdaIGN by developing two selection policies, i.e., NSP and ESP, which learns a selection pattern for nodes and edges in a multimodal heterogeneous graph. Then a directed graph based fusion is adopted to  prevent the influence of current utterance by future ones. 

The main difference between these works and our proposed approach are two-folds. Firstly, existing works usually focus on fusing multimodal information in a symmetric framework which  assumes different modalities have equivalent quality. Due to the varying quality of different modalities in real-world scenarios, we argue that it is more reasonable to conduct multimodal fusion in an asymmetric manner. To the end, we propose a novel cross-modality augmented transformer (CMA-Transformer) by treating one modality as the central modality and the others as auxiliary ones. Secondly,  different from existing works, we propose to learn modality representations with different granularities and develop a novel hierarchical distillation strategy to boost the effectiveness of the semantic distribution integration by maintaining the consistency between them.

\section{Preliminaries} 
In this section, we introduce the task definition and unimodal feature extraction.

\subsection{Task definition}
In the task of multimodal emotion recognition in conversation, a conversation is defined as  a sequence of utterances $U=\{{u_1},{u_2}, \cdots, {u_n}\}$ uttered by $m$ speakers, where $n$ is the number of utterances.
The $i$-th utterance representation  $\textbf{u}_i$ is represented by three different modalities denoted as $\textbf{u}_i = \{\textbf{u}_i^t,\textbf{u}_i^a,\textbf{u}_i^v\}$, where $\textbf{u}_i^t \in \mathbb{R}^{d_t}$,  $\textbf{u}_i^a \in \mathbb{R}^{d_a}$, $\textbf{u}_i^v \in \mathbb{R}^{d_v}$ are the corresponding representations of textual, audio, visual modality, respectively. $d_t$, $d_a$ and $d_v$ are the dimentions of the three modalities. 
The corresponding speaker of each utterance $u_i$ is denoted by $s_{\phi(u_i)}$ where $\phi(u_i)$ indicates the speaker index of $u_i$. 
The emotion label of the $i$-th utterance $u_i$ is $y_i \in C$, where $C$ is the set of the emotion labels. 
The goal of the task is to predict the emotion label $y_i$ for a given utterance $u_i$ based on the multimodal information of utterances as well as the corresponding speaker signals in the conversation .

\begin{figure*}
  \centering
  \includegraphics[width=\textwidth]{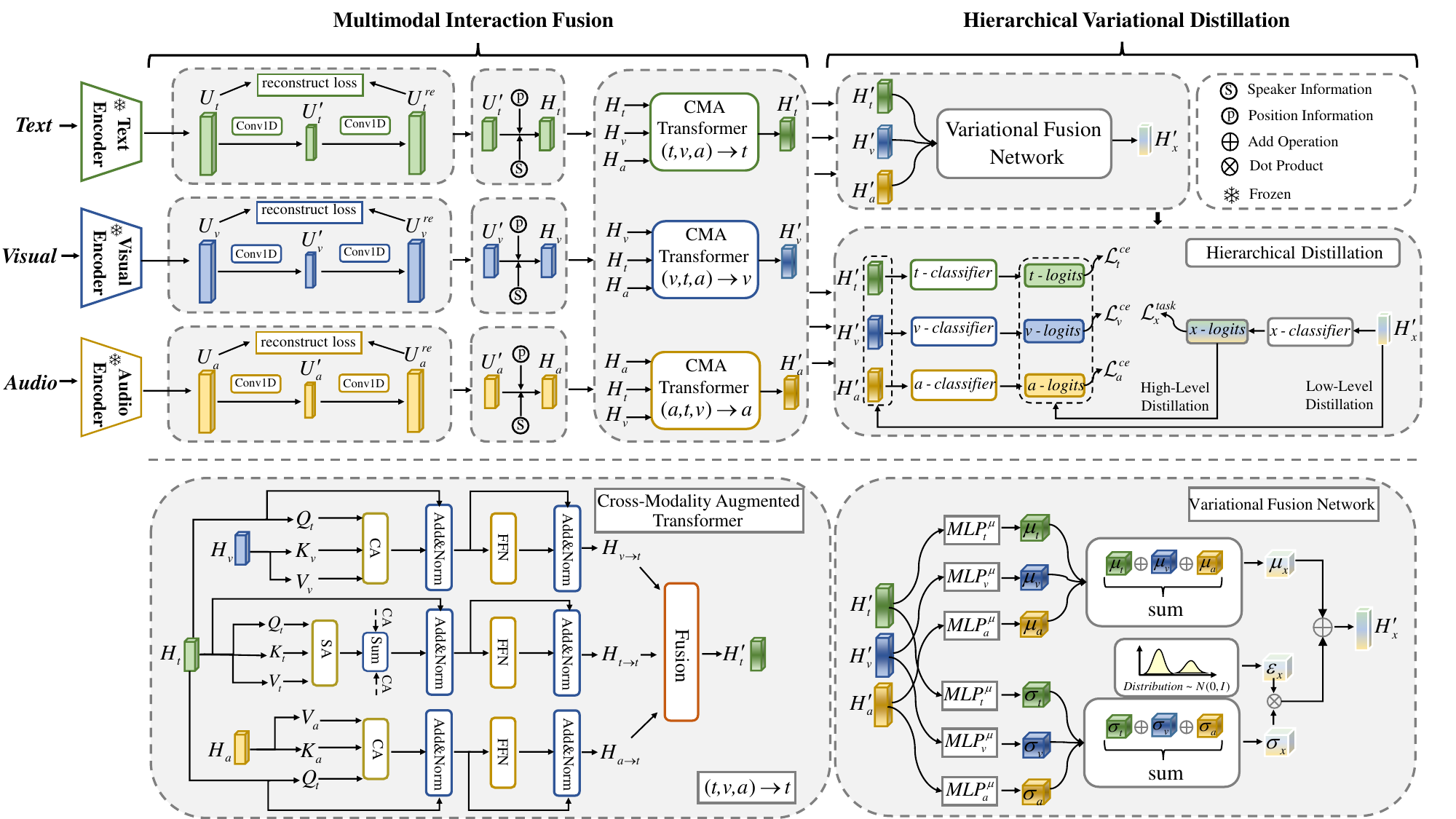}
  \caption{Overview of our proposed model CMATH, which consists of two modules: (1) Multimodal Interaction Fusion, which uses interaction information to improve modality quality; (2) Hierarchical Variational Distillation, which further boosts the fusion effectiveness between modalities.}
  \label{fig: model}
\end{figure*}

\subsection{Unimodal Feature Extraction}
Pre-trained models are typically trained on large-scale datasets, which enables them to capture rich semantic information. Therefore, we utilize pre-trained models to extract feature representations for each modality:

\textbf{Textual Modality:} we utilize the RoBERTa-Large \citep{roberta}  for extracting textual features. RoBERTa is an enhancement of the BERT, which utilizes a multi-layer transformer  to learn textual representations. For feature representation, we adopt the embeddings of the [CLS] tokens in the final layer of RoBERTa, which is a feature vector with a dimensionality of 1024.

\textbf{Audio Modality:} we utilize the OpenSMlLE Toolkit \citep{opensmile} to extract acoustic features. OpenSMlLE serves as a versatile feature extraction framework designed for signal processing, which allows for the configuration of modular feature extraction components via a scriptable console application. After processing with OpenSMlLE, the dimensionality of the acoustic feature representation is reduced to 1582 for the IEMOCAP dataset \citep{iemocap} and 300 for the MELD dataset \citep{meld}.

\textbf{Visual Modality:} we utilize the DenseNet \citep{forrest2014densenet} for the extraction of visual features. DenseNet is a convolutional neural network (CNN) that is known for its high efficiency. DenseNet outputs a feature representation with a dimensionality of 342.

\section{Methodology} 
In this section, we introduce the details of CMATH which consists of two major components, including Multimodal Interaction Fusion module (in Section \ref{sec:mif}) and Hierarchical Variational Distillation module (in Section \ref{sec:distillation}). The architecture of our proposed model is shown in Figure \ref{fig: model}.

\subsection{Multimodal Interaction Fusion}
\label{sec:mif}
The multimodal interaction fusion module consists of two submodules, i.e., Modality Reconstruction and Cross-Modality Augmented Transformer (CMA-Transformer).

\subsubsection{Modality Reconstruction}

To ensure the representations of different modalities lie in the same space as well as achieve high-quality compressed representation of each modality, we introduce the modality reconstruction submodule, which is  a simplified version of U-Net \citep{u-net}. Specifically, we adopt a 1D convolutional layer for each modality  to conduct down-sampling operation  and obtain their corresponding compressed representations in the common space. Then another 1D convolutional layer is leveraged for each modality to conduct up-sampling operation and reconstruct the  input representation. Formally, we have:
\begin{align} 
{\textbf{U}'_m} &= Conv1D'({\textbf{U}_m},{k_m}),m \in \{ t,a,v\}, \hfill \\
\textbf{U}_m^{''} &= Conv1D''({\textbf{U}'_m},{k_m}),m \in \{ t,a,v\}, \hfill 
\end{align}
where $\textbf{U}'_m \in {\mathbb{R}^{n \times d}}$, $\textbf{U}_m^{''} \in {\mathbb{R}^{n \times d_m}}$, $\textbf{U}_m$ is the raw utterance representation of the modality $m$, $k_m$ is the kernel size. $d$ and $d_m$ are the corresponding dimension size of the common space and the modality $m$. The reconstruction loss is defined as:
\begin{align}
{\cal L}_m^{re} &=\left\| \textbf{U}_m - \textbf{U}_m^{''} \right\|_F^2,\hfill \\ 
{\mathcal{L}^{re}} &= \sum\limits_{m \in \{ t,a,v\} } {\mathcal{L}_m^{re}}, \hfill    
\end{align} 
where $\left\| \cdot \right\|_F^2$ is the squared Frobenius norm.

Since speaker and positional information are crucial for emotion recognition in conversation \citep{speaker}, we integrate them  into  modality features.  To be specific, for each speaker $s_j, j \in \{1,\cdots, m\}$, we obtain its corresponding representation $\textbf{s}_j\in {\mathbb{R}^d}$ as follows:
\begin{align}
\textbf{s}_j = {\textbf{V}_s} \cdot \textbf{o}({s_j}),j = 1, 2, \cdots ,m
\end{align}
where ${\textbf{V}_s} \in {\mathbb{R}^{d \times m}}$ is the trainable embedding matrix, and $\textbf{o}({s_j}) \in {\mathbb{R}^m}$ is the one-hot vector of speaker $s_j$. The speaker embedding in the conversation is denoted as $\textbf{S} = \{\textbf{s}_{\phi({u_1})}, \textbf{s}_{\phi({u_2})}, \cdots,\textbf{s}_{\phi({u_n})}\} \in {\mathbb{R}^{n \times d}}$. 

For positional information, we employ sinusoidal positional encoding to encode positions within the conversation, which is defined as follows:
\begin{align} 
\textbf{p}_i^{2k} &= sin(\frac{i}{10000^{2k/d}}),\\
\textbf{p}_i^{2k+1} &= cos(\frac{i}{10000^{2k/d}}),
\end{align}
where $\textbf{P} = \{\textbf{p}_1, \cdots, \textbf{p}_n\} \in {\mathbb{R}^{n \times d}}$,  $i$ is the positional index of an utterance and $k$ is the dimension.

Therefore, we obtain the sequence representation of modality $m$ by adding 
the latent embedding $\textbf{U}'_m$, the speaker embedding $\textbf{S}$ and the positional embedding $\textbf{P}$:
\begin{equation}
    {\textbf{H}_m} = {\textbf{U}'_m}  + \textbf{S} + \textbf{P},
\end{equation}
where $\textbf{H}_m \in \mathbb{R}^{n \times d}$ is the augmented representation for modality $m$.

\subsubsection{CMA-Transformer}
Previous works usually assume different modalities have equivalent quality and 
focus on fusing multi-model information based on a symmetric architecture \citep{lian2021ctnet, yang2023scmm, mao2020dialoguetrm}. However, the qualities of different modalities would be varying and resulting in 
a suboptimal performance of these models. To the end, we develop an asymmetric fusion strategy named Cross-Modality Augmented Transformer (CMA-Transformer), which treats one modality as the central modality and takes others as auxiliary modalities. CMA-Transformer aims to enhance the representation of the central modality by exploiting information from these auxiliary modalities.
Since the textual modality derived CMA-Transformer have the same structure with the two other modalities (i.e., audio and visual) derived  CMA-Transformers, we mainly focus on explaining the  CMA-Transformer framework centered on the textual modality. 

For the textual modality, we treat it as the central-modality and enhance its representation by exploring signals from the audio modality (i.e., auxiliary modality):  
\begin{equation}\label{eq:a2t}
\begin{gathered}
  {{\tilde{\textbf{H}}}_{a \to t}} = CA({\textbf{H}_t},{\textbf{H}_a},{\textbf{H}_a}), \hfill \\
  {{\bar{\textbf{H}}}_{a \to t}} = Norm({{\tilde{\textbf{H}}}_{a \to t}} + {\textbf{H}_t}), \hfill \\
  {\textbf{H}_{a \to t}} = Norm(FFN({{\bar{\textbf{H}}}_{a \to t}}) + {{\bar{\textbf{H}}}}_{a \to t}),
\end{gathered}
\end{equation}
where ${{\tilde{\textbf{H}}}_{a \to t}}, {{\bar{\textbf{H}}}_{a \to t}}, {\textbf{H}_{a \to t}} \in \mathbb{R}^{n\times d}$, $CA(\cdot)$, $Norm(\cdot)$ and $FFN(\cdot)$ indicate cross-attention, normalization and  feed-forward network, respectively.

Similarly, we also explore signals from the visual modality (i.e., auxiliary modality) to boost the representation of the  textual modality: 
\begin{equation}\label{eq:v2t}
\begin{gathered}
  {{\tilde{\textbf{H}}}_{v \to t}} = CA({\textbf{H}_t},{\textbf{H}_v},{\textbf{H}_v}), \hfill \\
  {{\bar{\textbf{H}}}_{v \to t}} = Norm({{\tilde{\textbf{H}}}_{v \to t}} + {\textbf{H}_t}), \hfill \\
  {\textbf{H}_{v \to t}} = Norm(FFN({{\bar{\textbf{H}}}_{v \to t}}) + {{\bar{\textbf{H}}}_{v \to t}}),
\end{gathered}
\end{equation}
where ${{\tilde{\textbf{H}}}_{v \to t}}, {{\bar{\textbf{H}}}_{v \to t}}, {\textbf{H}_{v \to t}} \in \mathbb{R}^{n \times d}$.

Since the textual modality is the central modality, we first strengthen its representation via self-attention \citep{vaswani2017attention}, and then inject information from other two modalities into the textual modality: 
\begin{equation}\label{eq:t2t}
\begin{gathered}    
    {\tilde{\textbf{H}}_{t \to t}} = SA({\textbf{H}_t},{\textbf{H}_t},{\textbf{H}_t}), \hfill \\
  {{\tilde{\textbf{H}}}_{t \to t}} = {{\tilde{\textbf{H}}}_{t \to t}} + {{\tilde{\textbf{H}}}_{a \to t}} + {{\tilde{\textbf{H}}}_{v \to t}}, \hfill \\
  {{\bar{\textbf{H}}}_{t \to t}} = Norm({{\tilde{\textbf{H}}}_{t \to t}} + {\textbf{H}_t}), \hfill \\
  {\textbf{H}_{t \to t}} = Norm(FFN({{\bar{\textbf{H}}}_{t \to t}}) + {{\bar{\textbf{H}}}_{t \to t}}), \hfill 
\end{gathered}
\end{equation} 
where ${{\tilde{\textbf{H}}}_{t \to t}}, {{\bar{\textbf{H}}}_{t \to t}}, {\textbf{H}_{t \to t}} \in \mathbb{R}^{n \times d}$.

Finally, we adopt the gating fusion mechanism  to adaptively fuse these representations, making the final representation of the textual modality:
\begin{equation} \label{eq:gate}
    {\textbf{H}'_t} = Gate({\textbf{H}_{t \to t}},{\textbf{H}_{a \to t}},{\textbf{H}_{v \to t}}),
\end{equation}
We represent the process of the CMA-Transformer framework centered on the textual modality, i.e., Equations (\ref{eq:a2t}-\ref{eq:gate}), in a unified form:
\begin{equation}
    {\textbf{H}'_t} = {\text{CMA-Transformer}_{\theta_t}}({\textbf{H}_t},{\textbf{H}_a},{\textbf{H}_v}),
\end{equation}
where $\textbf{H}'_t \in \mathbb{R}^{n \times d}$, and $\theta_t$ are trainable parameters.

Similarly, we can also obtain the enhanced representations centered on audio and visual modalities as follows:
\begin{equation}
\begin{gathered}
  {\textbf{H}'_a} = {\text{CMA-Transformer}_{\theta_a}}({\textbf{H}_a},{\textbf{H}_t},{\textbf{H}_v}), \hfill \\
  {\textbf{H}'_v} = {\text{CMA-Transformer}_{\theta_v}}({\textbf{H}_v},{\textbf{H}_t},{\textbf{H}_a}), \hfill \\ 
\end{gathered} 
\end{equation}
where $\textbf{H}'_a \in \mathbb{R}^{n \times d}$ and $\textbf{H}'_v \in \mathbb{R}^{n \times d}$. 

\subsection{Hierarchical Variational Distillation}
\label{sec:distillation}
In this section, we introduce the Hierarchical Variational Distillation module, which consists of two submodules, i.e., the Variational Fusion Network and the Hierarchical Distillation.

\subsubsection{Variational Fusion Network}
\label{sec:vfs}
After we obtain the augmented representations for each modality based on CMA transformer, we first apply a variational fusion network to fuse these fine-grained representations for each modality into a coarse-grained representations.
To be specific,  we assume the latent representations of each modality are represented by a Gaussian distribution ${\mathcal{N}_m}({\boldsymbol{\mu} _m},{\boldsymbol{\sigma} ^2_m})$, which takes $\textbf{H}'_m$ as input and outputs its corresponding mean $\boldsymbol{\mu} _m\in {\mathbb{R}^{n \times d}}$  and standard deviation $\boldsymbol{\sigma} ^2_m\in {\mathbb{R}^{n \times d}}$ of each modality:
\begin{equation}
\begin{gathered}
  {\boldsymbol{\mu} _m} = MLP_m^\mu ({\textbf{H}'_m}), \hfill \\
  {\boldsymbol{\sigma} _m} = MLP_m^\sigma ({\textbf{H}'_m}), \hfill \\ 
\end{gathered} 
\end{equation}
where ${MLP_m^\mu}(\cdot) $, ${MLP_m^\sigma}(\cdot) $ represent two multi-layer perceptrons.

Then, we assume these Gaussian distributions of different modalities  are independent and integrate them into a multimodal  Gaussian distribution ${\mathcal{N}_x}({\boldsymbol{\mu} _x},{\boldsymbol{\sigma} ^2_x})$. Specifically, we obtain ${\boldsymbol{\mu} _x \in \mathbb{R}^{n \times d}}$ and ${\boldsymbol{\sigma} _x \in \mathbb{R}^{n \times  d}}$ by 
simply averaging the means and standard deviations of all Gaussian distributions:
\begin{equation}
\begin{gathered}
  {\boldsymbol{\mu} _x} =  \frac{1}{|m|}\sum\limits_{m \in \{ t,a,v\} } {{\boldsymbol{\mu} _m}} , \hfill \\
  {\boldsymbol{\sigma} _x} = \frac{1}{|m|}\sum\limits_{m \in \{ t,a,v\} } {{\boldsymbol{\sigma} _m}} . \hfill \\ 
\end{gathered} 
\end{equation}
The representation sampled from the learned multimodal Gaussian distribution can be regarded as the fused representations of all modalities.
Since the sampling operation is not differentiable, we adopt the reparameterization trick \citep{kingma2014auto} for allowing back-propagation of gradients during the training stage. Specifically, the fused multimodal representations ${\textbf{H}'_x} \in \mathbb{R}^{n \times d}$ is generated as follows:
\begin{equation}
    {\textbf{H}'_x} = {\boldsymbol{\mu} _x} + \epsilon  \cdot {\boldsymbol{\sigma} _x},
\end{equation}
where $\epsilon$ is the random noise sampled from a normal distribution $\mathcal{N}(0,\textbf{I})$. 

\subsubsection{Hierarchical Distillation}
\label{sec:hds}
With the fine-grained representations $\textbf{H}'_m$ learned by CMA-Transformer for each modality $m$ and the coarse-grained representation $\textbf{H}'_x$ learned by the variational fusion network on hand, we introduce the hierarchical distillation submodule to maintain the consistency between modality representations with different granularities. To be specific, the hierarchical distillation paradigm is comprised of a low-level distillation and a high-level distillation. The former is designed to minimize the difference between semantic logits of teacher $\textbf{H}'_x$ and student $\textbf{H}'_m$. The latter is developed to minimize the KL-Divergence between decision logits of teacher $\hat{Y}_x$ and student teacher $\hat{Y}_m$. 

\textbf{Low-Level Distillation:}  the objective of the low-level distillation is to impose consistency constraints between modalities on the semantic representation spaces. Specifically, we treat ${\textbf{H}'_x}$ as teacher and ${\textbf{H}'_m}$ as student,  and adopt  Mean Squared Error (MSE) loss \citep{willmott2005mse} as the low-level distillation loss:
\begin{equation}
\begin{gathered}
    \mathcal{L}_m^{mse} =  \left\| {{\textbf{H}'_x} - {{\textbf{H}'}_m}} \right\|_F^2, \hfill \\ 
    {\mathcal{L}^{mse}} = \sum\limits_{m \in \{ t,a,v\} } {\mathcal{L}_m^{mse}}, \hfill \\
\end{gathered}
\end{equation}
where $\left\| \cdot \right\|_F^2$ is the squared Frobenius norm.

\textbf{High-Level Distillation:} the objective of high-level distillation is to maintain consistency  between modalities on the decision representation spaces. 
We first obtain the classification results of modalities based on different granularities:
\begin{equation}
    \begin{gathered}
  {{\hat Y}_m} = \text{Classifier}_m({\textbf{H}'_m}), \hfill \\
  {{\hat Y}_x} = \text{Classifier}_x({\textbf{H}'_x}), \hfill \\
\end{gathered} 
\end{equation}
where $\text{Classifier}_m(\cdot)$ and $\text{Classifier}_x(\cdot)$ are composed of a fully connected layer followed by a softmax layer.
We utilize the cross-entropy loss for training these classifiers. In particular,  the cross-entropy loss between the predicted label distribution $\hat{Y}_m$ and the true label distribution $Y$ is defined as follows: 
\begin{equation} 
    \mathcal{L}_m^{ce} =  - \frac{1}{n}\sum\limits_{i = 1}^n {\sum\limits_{j = 1}^C {{y_{i,j}}\log ({{\hat y}^m_{i,j}})} },  
\end{equation}
where ${\hat y}^m_i \in \hat{Y}_m$ is the predicted label for the $i$-th utterance $u_i$ based on the fine-grained representations ${\textbf{H}'_m}$, and $y_i \in Y$ is the true label for the $i$-th utterance $u_i$.
Similarly, we define the cross-entropy loss between the predicted label distribution $\hat{Y}_x$ and the true label distribution $Y$ as follows: 
\begin{equation}
    \mathcal{L}_x^{ce} =  - \frac{1}{n}\sum\limits_{i = 1}^n {\sum\limits_{j = 1}^C {{y_{i,j}}\log ({{\hat y}^x_{i,j}})} },
\end{equation}
where ${\hat y}^x_i \in \hat{Y}_x$ is the predicted label for the $i$-th utterance $u_i$ based on the coarse-grained representations ${\textbf{H}'_x}$.
The overall cross-entropy loss is defined as:
\begin{equation}  
    {\mathcal{L}^{ce}} = \mathcal{L}_x^{ce} + \sum\limits_{m \in \{ t,a,v\} } {\mathcal{L}_m^{ce}}. 
\end{equation}

Then, we adopt  KL-Divergence \citep{hershey2007kl} loss as the high-level distillation loss, where $\hat{Y}_x$ and $\hat{Y}_m$ are treated as teacher and student, respectively. Formally, we have:
\begin{equation}
\begin{gathered}
  \mathcal{L}_m^{kl} = \frac{1}{n}\sum\limits_{i = 1}^n {\sum\limits_{j = 1}^C {{{\hat y}^x_{i,j}}\log (\frac{{{{\hat y}^x_{i,j}}}}{{{{\hat y}^m_{i,j}}}})} },  \hfill \\ 
  {\mathcal{L}^{kl}} = \sum\limits_{m \in \{ t,a,v\} } {\mathcal{L}_m^{kl}}. \hfill \\
\end{gathered} 
\end{equation}
Finally, the overall loss is defined as follows:
\begin{equation}
    \begin{gathered}
     {\mathcal{L}_{all}} = \mathcal{L}^{ce} + \mathcal{L}^{re} + {\gamma _1}{\mathcal{L}^{mse}} + {\gamma _2}{\mathcal{L}^{kl}} , \hfill \\
\end{gathered} 
\end{equation}
where $\gamma_1$ and $\gamma_2$  are hyperparameters.

\section{Experiments}
\subsection{Datasets and Evaluations}
We conduct experiments on two well-known benchmark datasets, i.e.,  IEMOCAP \citep{iemocap} and MELD \citep{meld}, and the statistics of  these datasets are detailed in Table \ref{tab:dataset}.
\newline
\indent
\textbf{IEMOCAP}. This dataset comprises two-way conversations involving ten speakers, with a total of 151 conversations and 7,433 utterances. It is segmented into five sessions, with the first four utilized for training and the final session reserved for testing. Following \citep{ma2022mvn}, 
we randomly select 20\% of the training set as
the validation set.  
Each utterance within the dataset is annotated with one of six emotions, including \textit{Happy, Sad, Neutral, Angry, Excited, Frustrated. }
\newline
\indent
\textbf{MELD}. Different from IEMOCAP, MELD is a multi-speaker conversation dataset with three or more speakers in a conversation.  It is collected from the TV series \textit{Friends}, which includes 
1,433 conversations and 13,708 utterances. Each utterance is categorized under one of the seven emotions, i.e., \textit{Neutral, Surprise, Fear, Sadness, Joy, Disgust, Angry}.
\newline
\indent
\textbf{Evaluation Metrics}. We evaluate the model's performance based on two metrics, i.e.,  the overall accuracy (ACC) and the weighted average F1-score (W-F1). Additionally, we provide the F1-score for each emotion category to offer a comprehensive assessment of performance.


\begin{table*}[]
\centering
\caption{Statistics of IEMOCAP and MELD datasets.}
\begin{tabular}{cccccccccc}
\toprule
\multirow{2}{*}{Dataset} & \multicolumn{2}{c}{Conversations} & \multicolumn{2}{c}{Utterances} & \multicolumn{2}{c}{Utterances per   Conversation} & Classes &  &  \\
\cline{2-10}
                         & Train+Val         & Test         & Train+Val                 & Test       & Train+Val  & Test \\
\hline
IEMOCAP                  & 120               & 31           & 5810             & 1623       & 48.42                     & 52.35                 & 6       &  &  \\
MELD                     & 1153              & 280          & 11098            & 2610       & 9.62                      & 9.32                  & 7       &  &  \\
\bottomrule
\end{tabular}
\label{tab:dataset}
\end{table*}

\subsection{Baselines}
We conduct a comprehensive comparison of our proposed model with various SOTA baseline methods:
\begin{itemize}
\item \textbf{DialogueRNN}\citep{majumder2019dialoguernn}: It models the speaker, the preceding context and the preceding emotion via three gated recurrent units, including a global GRU, a party GRU and an emotion GRU. 
\item \textbf{DialogueGCN}\citep{ghosal2019dialoguegcn}: This baseline considers the inter-speaker dependency  and the intra-speaker dependency to capture emotional influence between different speakers and emotional inertia of individual speakers, respectively. The graph convolution network is adopted to model contextual information among distant utterances.
\item \textbf{MMGCN}\citep{hu2021mmgcn}: MMGCN applies a multimodal fused graph convolutional network to explore the multimodal information and capture distant contextual information. It can effectively model multimodal dependencies  as well as inter- and intra-speaker dependencies.
\item \textbf{CTNet}\citep{lian2021ctnet}: It utilizes transformer-based structures to model intra-modal and cross-modal interactions of different modalities. A single-modal transformer is developed to capture temporal dependencies among unimodal features, and a cross-modal transformer is designed to learn cross-modal interactions on the unaligned multimodal features. 
\item \textbf{MM-DFN}\citep{hu2022mmdfn}: It attempts to reduce redundancy and enhance complementarity between modalities by designing a graph-based dynamic fusion module to merge multimodal context information.  It employs the graph convolution operation  to aggregate intra- and inter-modality contextual information in a specific space, and adopts the gating mechanism to learn intrinsic sequential patterns of contextual information in adjacent semantic spaces. 
\item \textbf{SCMM}\citep{yang2023scmm}: It aims to model the varying difficulty of each utterance and the specific contribution of each modality. A self-adaptive path selection strategy is adopted to select an appropriate path to obtain utterance representation. 
\item \textbf{CMCF-SRNet}\citep{zhang2023cmcf-srnet}: It designs two transformers, i.e., cross-modal locality-constrained transformer and graph-based semantic refinement transformer, to explore the multimodal interaction and semantic relationship information among utterances. 
\item \textbf{MultiDAG}\citep{nguyen2024curriculum}: It applies directed acyclic graph to fuse multimodal features within a unified framework. Then the   curriculum learning is leveraged  to facilitate the learning process by gradually presenting training samples in a meaningful order to address emotional shift issues and imbalanced data.
\item \textbf{AdaIGN}\citep{tu2024adaptive}: It presents an adaptive interactive graph network to balance the intra- and inter-speaker emotion dependency  as well as mitigate the issue of conflicting emotion across different modalities.   It further adopts a directed graph based fusion strategy to prevent the influence of current utterance by future ones. 
\end{itemize}

\subsection{Implementation Details}
We implement the proposed model using Pytorch on NVIDIA RTX 2080Ti and adopt  Adam \citep{adam} as the optimizer with an initial learning rate of 3.0e-3 for IEMOCAP and 2.0e-4 for MELD. The batch size is 16 for IEMOCAP and 4 for MELD. We set $\gamma_1$ and $\gamma_2$ to 1.0 and 1.8 for IEMOCAP, 1.0 and 1.2 for MELD. Note that  the two baselines DialogueRNN and DialogueGCN are initially developed  for unimodel scenario, we apply an early concatenation operation for fusing multimodal features in their implementation. All results are averages of 5 runs.

\begin{table*}[]
\centering
\caption{Overall performance comparison   on IEMOCAP dataset. The best and second best are in bold and underlined, respectively. '$\dag$' means the improvement of CMATH over  the best performing baseline AdaIGN is significant at p<0.05 based on t-test. The metric for each emotion category is F1-score.}
\begin{tabular}{c|cccccc|cc}
\toprule
\multirow{2}{*}{Model} & \multicolumn{8}{c}{IEMOCAP}                                            
\\
\cline{2-9}
                       & Happy & Sad   & Neutral & Angry & Excited & Frustrated & ACC   & W-F1    \\
\midrule
DialogueRNN                 & 32.20  & 80.26  & 57.89   & 62.82  & 73.87   & 59.76      & 63.52  & 62.89   \\
DialogueGCN                 & 51.57  & 80.48  & 57.69   & 53.95  & 72.81   & 57.33      & 63.22  & 62.89   \\
MMGCN                       & 45.14  & 77.16  & 64.36   & 68.82 & 74.71   & 61.40      & 66.36 & 66.26   \\
CTNet                       & 51.30  & 79.90  & 65.80   & 67.20 & \underline{78.70}   & 58.80      & 68.00 & 67.50   \\
MM-DFN                      & 42.22  & 78.98  & 66.42   & 69.77 & 75.56   & 66.33      & 68.21  & 68.18   \\
SCMM                           & 45.37  & 78.76 & 63.54   & 66.05 & 76.70   & 66.18      & - & 67.53   \\
CMCF-SRNet                    & 52.20  & 80.90 & 68.80   & \underline{70.30} & 76.70   & 61.60      & \underline{70.50} & 69.60   \\
MultiDAG                   & 45.26  & \underline{81.40} & 69.53   & \textbf{70.33} & 71.61   & 66.94      & 69.11 & 69.08   \\
AdaIGN                        & 53.04  & \textbf{81.47} & \underline{71.26}   & 65.87 & 76.34   & \underline{67.79}      & 70.49 & \underline{70.74}   \\
\midrule
CMATH                       & $\textbf{65.63}^{\dag}$ & 79.68 & $\textbf{75.65}^{\dag}$   & 69.55 & $\textbf{80.72}^{\dag}$   &$\textbf{68.82}^{\dag}$      & $\textbf{73.90}^{\dag}$ & $\textbf{73.96}^{\dag}$  
\\
\bottomrule
\end{tabular}
\label{tab:iemocap}
\end{table*}

\begin{table*}[]
\centering
\caption{Overall performance comparison on MELD dataset. The best and second best are in bold and underlined, respectively. '$\dag$' means the improvement of CMATH over  the best performing baseline AdaIGN is significant at p<0.05 based on t-test. The metric for each emotion category is F1-score.}
\begin{tabular}{c|ccccccc|cc}
\toprule
\multirow{2}{*}{Model} & \multicolumn{9}{c}{MELD}                                                       \\
\cline{2-10}
                       & Neutral & Surprise & Fear  & Sadness & Joy   & Disgust & Angry & ACC   & W-F1  \\
\midrule
DialogueRNN                  & 76.97   & 47.69    & -  & 20.41   & 50.92  & -    & 45.52  & 60.31  & 57.66 \\
DialogueGCN                   & 75.97   & 46.05    & -  & 19.60   & 51.20  & -    & 40.83  & 58.62   & 56.36 \\
MMGCN                   & 76.33   & 48.15    & -   & 22.93   & 53.02  & -  & 46.09 & 60.42 & 58.31 \\
CTNet                  & 77.40   & 52.70    & \underline{10.00} & 32.50   & 56.00 & \underline{11.20}   & 44.60 & 62.00     & 60.50 \\
MM-DFN                  & 77.76   & 50.69    & -  & 22.93   & 54.78  & -   & 47.82 & 62.49 & 59.46 \\
SCMM                     & -   & -    & -   & -  & -  & -  & -  & -  & 59.44   \\
CMCF-SRNet                & -   & -    & -   & -  & -  & -  & -  & 62.80 &62.30   \\
MultiDAG                  & -   & -    & -   & -  & -  & -  & -  & 64.41 & 64.00 \\
AdaIGN                  & \underline{79.75}   & \textbf{60.53}    & -     & \textbf{43.70}   & \textbf{64.54} & -       & \underline{56.15} & \underline{67.62} & \underline{66.79} \\
\midrule
CMATH                       & $\textbf{80.04}^{\dag}$   & \underline{59.09}    & $\textbf{19.86}$ & \underline{43.57}   & \underline{64.09} & $\textbf{25.90}$   & $\textbf{57.83}^{\dag}$ & $\textbf{68.34}^{\dag}$&$\textbf{66.93}^{\dag}$ \\
\bottomrule
\end{tabular}
\label{tab:meld}
\end{table*}

\subsection{Overall Performances}
Table \ref{tab:iemocap} and table \ref{tab:meld} report the experimental results  of all baseline models and our proposed method on the IEMOCAP and MELD datasets, respectively. The best and second best results are bolded and underlined, respectively.
From the results, we can observe that both DialogueRNN and DialogueGCN obtain the worst performance. This is because they are initially designed for single modality. Although an early concatenation fusion is utilized in their implementations, the complex interaction information among different modalities are not well explored by them. 
Compared to DialogueRNN and DialogueGCN, these graph-based or transformer-based baselines demonstrate superior performances as they can better explore the complex multimodal interaction information. 
Among all baselines, AdaIGN obtains the best performance in most cases. Our proposed method is   better than all baselines.   
To be specific, on the IEMOCAP dataset, our model outperforms the best performing baseline AdaIGN by  3.41\% and 3.22\%  in terms of the metrics ACC and W-F1, respectively.  We further conduct significance test between our model and the best performing baseline, and the results suggest that the improvement of CMATH over AdaIGN is significant on both datasets with $p < 0.05$ based on t-test. In addition, we also analyze the detailed results on each  emotion category, and the results show that our model achieves superior or competitive performances on most emotion categories.

\subsection{Ablation Study}
\begin{table}[]
\centering
\caption{Ablation experiments for exploring the importance of each components. ${\dag}$ means that the improvement is statistically significant
(t-test with p < 0.05)}
\begin{tabular}{l|cccc}
\toprule
\multirow{2}{*}{} & \multicolumn{2}{c}{IEMCOAP} & \multicolumn{2}{c}{MELD} \\
\cline{2-5}
                  & ACC          & W-F1         & ACC         & W-F1       \\
\hline
w/o MR           & 72.49        & 72.67        & 67.64       & 66.60     \\
w/o CMA-T        & 68.15        & 68.48        & 67.86       & 66.69      \\
w/o LLD          & 73.60        & 73.76        & 68.12       & 66.86      \\
w/o HLD          & 71.05        & 71.18        & 68.13       & 66.75      \\
w/o HD           & 71.48        & 71.66        & 67.63       & 66.63      \\
\midrule
CMATH            & $\textbf{73.90}^{\dag}$        & $\textbf{73.96}^{\dag}$        & $\textbf{68.34}^{\dag}$       &$ \textbf{66.93}^{\dag}$     \\
\bottomrule
\end{tabular}
\label{tab:ab}
\end{table}

In this section, we conduct the ablation study  to investigate the contribution of each component in our proposed model. Specifically, we introduce the following variants of CMATH to perform the comparison: 
\begin{itemize}
\item \textbf{w/o MR}:  We replace the modality reconstruction module by a fully-connected layer, and still keep the speaker and positional information.
\item \textbf{w/o CMA-T}:  We ignore the asymmetric fusion operation by
discarding the CMA Transformer.
\item \textbf{w/o LLD}: We remove the low-level distillation part, and only consider the high-level distillation. 
\item \textbf{w/o HLD}: We discard the high-level distillation part, and only keep the low-level distillation.
\item \textbf{w/o HD}: We ignore the consistency operation by discarding the hierarchical distillation.
\end{itemize}

The experimental results  are shown in Table \ref{tab:ab} and we can observe that CMATH achieves the best performance compared to each variant, which indicates the rationality of its design. Specifically,  the modality reconstruction module helps to improve the performance by obtaining high-quality compressed representation of each modality. Moreover, there is a significant performance decline if we remove the CMA-Transformer module, which demonstrates the effectiveness of incorporating the asymmetric fusion strategy. The performance drops on the IEMOCAP dataset is larger than on the MELD dataset. The reason may be attributed to the conversation length in the IEMOCAP dataset is considerably longer than those in the MELD dataset (as shown in Table \ref{tab:dataset}), and CMA-Transformer is able to explore more interaction signals on IEMOCAP than on MELD. 
In addition, removing either the low-level distillation or the high-level distillation  will result in performance degradation, which demonstrates that
each distillation constraint plays a critical role in our model.
Besides, removing both low- and high-level distillation parts will further lead to  performance degradation  compared to removing each of them. It indicates that the low- and high-level distillation parts can compensates each other, and 
maintaining the consistency between modality representations with different granularities are crucial for the proposed model.

\subsection{Effect of Different Modality Settings}
\begin{table}[]
\centering
\caption{
Experiments for exploring the effect of a unique modality as well as a combination of multiple modalities.}
\begin{tabular}{c|cccc}
\toprule
\multirow{2}{*}{} & \multicolumn{2}{c}{IEMCOAP} & \multicolumn{2}{c}{MELD} \\
\cline{2-5}
                  & ACC          & W-F1         & ACC         & W-F1       \\
\hline
Text              & 67.98        & 67.81        & 67.26       & 65.85      \\
Audio             & 50.57        & 52.9         & 49.19       & 41.08      \\
Visual            & 27.73        & 30.02        & 48.12       & 31.27      \\
Text+Audio        & 70.55        & 70.58        & 67.93       & 66.56      \\
Text+Visual       & 68.45        & 68.37        & 67.89       & 66.63      \\
Audio+Visual      & 61.92        & 61.84        & 47.43       & 42.04      \\
\midrule
CMATH           & \textbf{73.90}        & \textbf{73.96}        & \textbf{68.34}       & \textbf{66.93}     \\
\bottomrule
\end{tabular}
\label{tab:modal-ab}
\end{table}

In order to evaluate the effectiveness of our model under different modality settings, we remove one or two modalities and the results are shown in Table \ref{tab:modal-ab}. We have the following observations. First, the performance of CMATH over the textual modality  is significantly superior to the other two modalities according to unimodal results, which indicates that textual features play the most important role in the ERC task. Among all modalities, the performance on visual modality is the worst which reflects that its quality is relatively lower than other modalities.
Second, when we implement CMATH based on any combination of two modalities, we can observe that the corresponding performance is much better than  its counterpart solely on a unique modality. 
Third, the best performance is achieved when all three modalities are leveraged simultaneously. This result confirms that our model can reasonably integrating information from different modalities.


\begin{figure}
  \centering
  \includegraphics[width=0.5\textwidth]{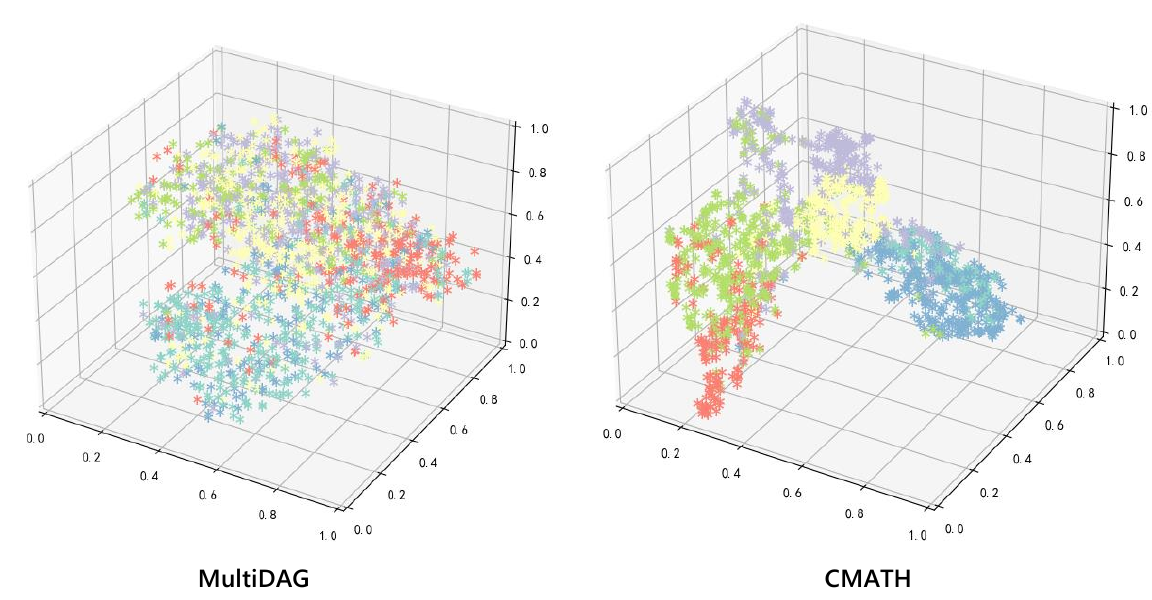}
  \caption{The t-SNE visualizations of learned embeddings by CMATH and MultiDAG on the IEMOCAP dataset.}
  \label{fig: tsne}
\end{figure}

\begin{figure}
  \centering
  \includegraphics[width=0.5\textwidth]{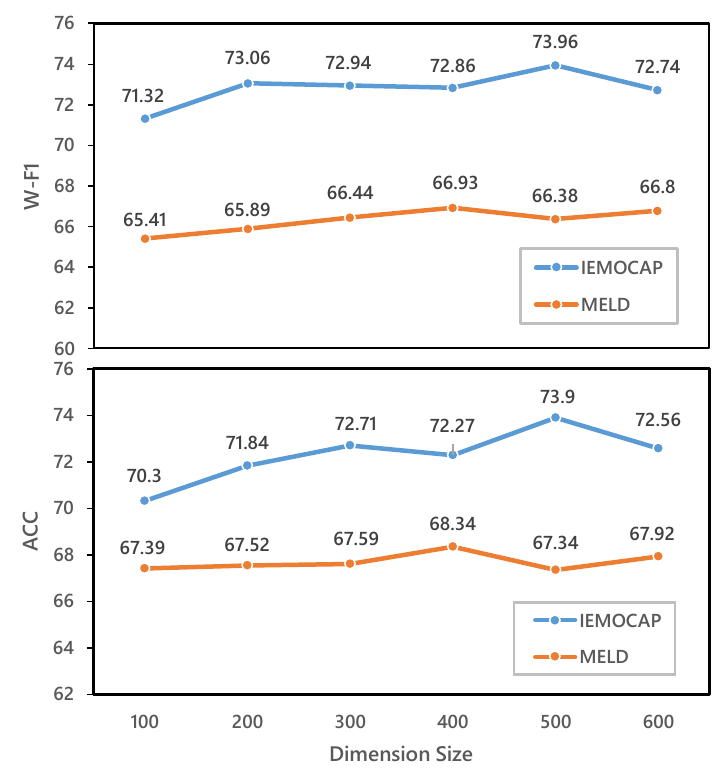}
  \caption{Performance of the proposed model with different dimension size on the two datasets.}
  \label{fig: dimension}
\end{figure}

\begin{figure}
  \centering
  \includegraphics[width=0.45\textwidth]{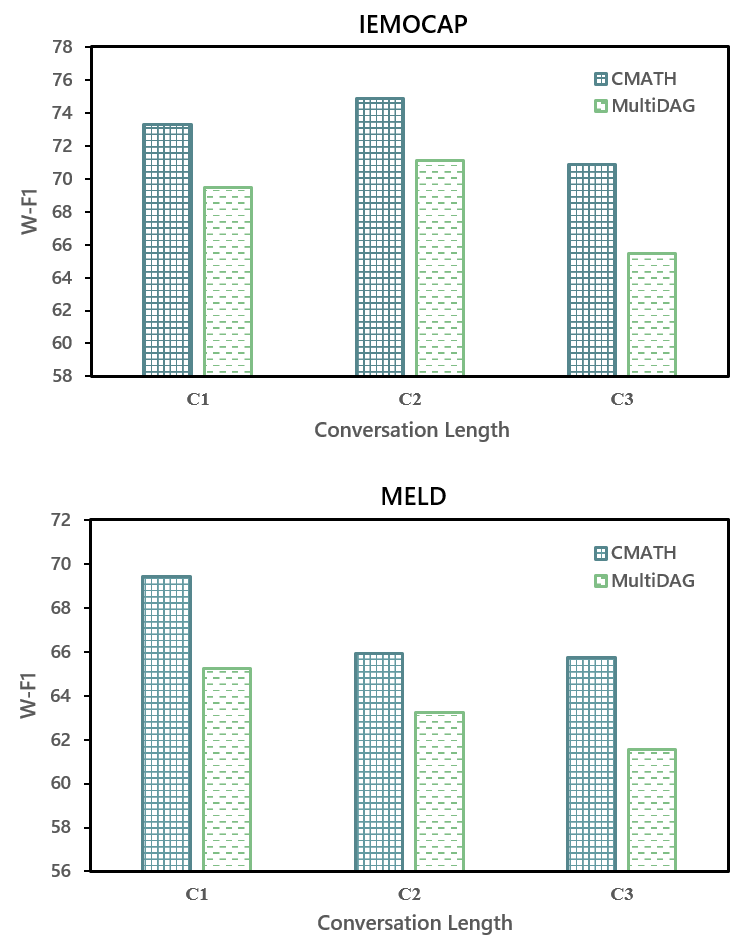}
  \caption{Performance of the proposed model with different conversation lengths on the two datasets.}
  \label{fig: length}
\end{figure}

\begin{figure*}
  \centering
  \includegraphics[width=0.9\textwidth]{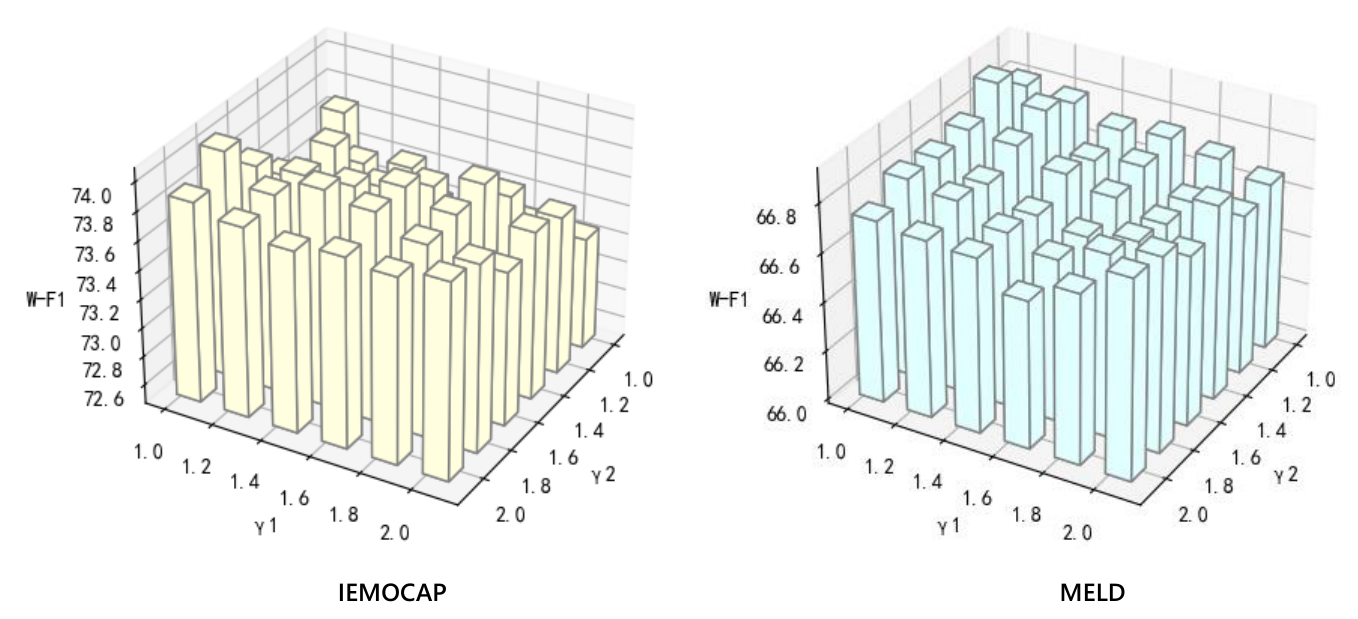}
  \caption{Parameter analysis of $\gamma_1$ and $\gamma_2$ on the two datasets.}
  \label{fig: para}
\end{figure*}

\begin{figure*}
  \centering
  \includegraphics[width=0.9\textwidth]{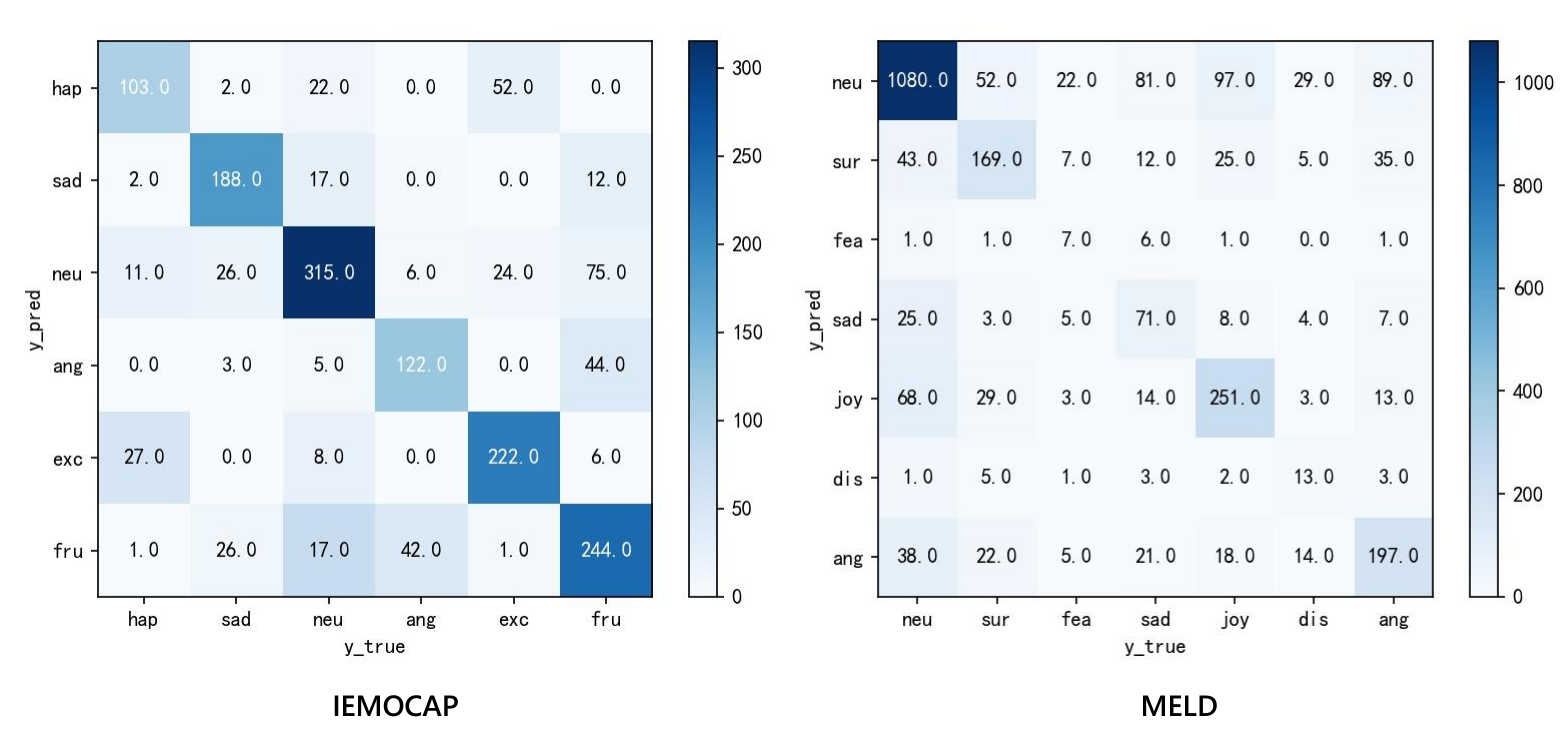}
  \caption{Confusion matrices for the proposed model on the two datasets.}
  \label{fig: confusion}
\end{figure*}

\subsection{Representation Visualization}
To further verify the superiority of our proposed model CMATH, we adopt t-SNE to illustrate learned embeddings by CMATH and MultiDAG\footnote{We do not compare with AdaIGN as its code is not available.} on the IEMOCAP dataset. Figure \ref{fig: tsne} presents the results, from which we can  find that the embeddings learned MultiDAG are not satisfactory as many nodes with different labels are mixed together. In contrast, our method is clearly superior to MultiDAG as the  learned embeddings are more distinguishable, which demonstrates a higher intra-class tightness and clear inter-class boundaries. The results validates the effectiveness of our proposed method.

\subsection{Impact of Embedding Dimension} 
In this section, we explore how the size of the embedding dimension affects the performance of our proposed model. 
We vary it from 100 to 600 with a step size of 100. Figure \ref{fig: dimension} shows the performance of CMATH with respect to different dimension sizes. For IEMOCAP, in the beginning, the performance of CMATH improves gradually as a higher size of embedding dimension is adopted, and reaches to the peak when the dimension size equals to 500. After that, the performance  starts to decline as the dimension size continues to increase. For MELD,  we can observe a similar trend of performance, where the best performance is obtained when the dimension size equals to 400.
The results suggests that the proposed model can produce the state-of-the-art performance in a relatively lower-dimensional space.

\subsection{Impact of Conversation Length}

To investigate the performance of our proposed method with different conversation lengths, we group test conversations  into three different categories with respect to their lengths.  
Specifically, for IEMOCAP, the conversation categories are defined as C1 (1-39 utterances), C2 (40-69 utterances), and C3 (over 70 utterances). For MELD, the conversation categories are defined as C1 (1-9 utterances), C2 (10-20 utterances), and C3 (over 20 utterances).

Figure \ref{fig: length} presents the performance of CMATH and MultiDAG with respect to different conversation categories. 
We can observe that our proposed model CMATH consistently outperforms MultiDAG across all categories on both datasets. Specifically, for IEMOCAP,  the relative improvements of CMATH over MultiDAG on the three categories are 3.86\%, 3.74\% and 5.41\%, respectively.  For MELD, the relative performance improvements of CMATH over MultiDAG  are 4.15\%, 2.67\% and 4.19\% on the three categories. 
The experimental results verify that our proposed model achieves better and more stable performance for various conversation lengths.

\subsection{Hyperparameter Analysis}
In this section, we explore the effects of hyperparameters $\gamma_1$ and $\gamma_2$ on the model performance, where $\gamma_1$ and $\gamma_2$ control the low- and high-level distillation losses, respectively. 
A higher $\gamma_1$ ($\gamma_2$) signifies a higher intensity of  fine-grained (coarse-grained) information constraint between modalities. 
We perform grid search over combinations of $\gamma_1$ and $\gamma_2$ by varying them from 1.0 to 2.0 with a step of 0.2, and the results are depicted in Figure \ref{fig: para}. 
We observe that our proposed model achieves the best performance when $\gamma_1=1.0$ $(1.0)$ and $\gamma_2=1.8$ $(1.2)$  on the IEMOCAP (MELD) dataset
in terms of the metric W-F1. This result indicates that  imposing consistency constraints between modalities on the decision representation spaces. (i.e., high-level distillation) is  relatively more important compared to maintaining consistency between modalities on the semantic representation spaces. It also verifies the effectiveness of developing the hierarchical distillation mechanism in our model, which can maintain the consistency between modality representations with different granularities.

\subsection{Error Analysis}
In this section, we report the confusion matrix of CMATH in Figure \ref{fig: confusion}. For both datasets, we can observe a high concentration of items on the diagonal line which reflects the performance of our model over each class.  Moreover, we can also observe some misclassified samples for different classes. To be specific, on the IEMOCAP dataset, we observe that  there are 48 misclassified samples for the class ``Angry'', among which most of them (i.e., 42 samples) are misclassified to the class ``Frustrated''. This may be attributed to the high similarity between the two classes. Similar observations can be found for the pair of classes, i.e.,  ``Happy'' and ``Excited''.  
On the MELD dataset, we find that most of the misclassified samples for each classes fall into the class ``Neutral''. The  reason may be that the number of samples of the class ``Neutral'' is overwhelmingly larger than other classes which makes the model biased to this dominant class. 

\section{Conclusion}

In this paper, we proposed  a novel model named CMATH for  multimodal emotion recognition in conversation. We develop a cross-modality augmented transformer for fine-grained feature fusion, and then propose a variational fusion network for coarse-grained semantic information fusion. After that, we introduce a Hierarchical Distillation module, which consists of both low- and high-level distillations, to maintain the consistency between modality representations with different granularities. 
We conducted extensive experiments on two public datasets, and the results demonstrate that our proposed model is significantly superior to the state-of-the-art baselines.

\printcredits

\section*{Acknowledgement}
This work was supported by the National Natural Science Foundation of China [grant number 62472059]; the Natural Science Foundation of Chongqing, China [grant number CSTB2022NSCQ-MSX1672]; the Chongqing Talent Plan Project, China [grant number CSTC2024YCJH-BGZXM0022]; the Major Project of Science and Technology Research Program of Chongqing Education Commission of China [grant number KJZD-M202201102].

\bibliographystyle{cas-model2-names}

\end{document}